\documentclass{article}
\usepackage[margin=1in]{geometry} 
 \usepackage{setspace}
\usepackage{appendix}
\usepackage{amsmath}
\usepackage{amsthm}
\usepackage{amssymb}
\usepackage{float}
\usepackage{siunitx}
\usepackage{graphicx, color}
\usepackage[hypcap=false]{caption}
\usepackage{capt-of}
\usepackage{subcaption}
\usepackage{authblk}
\usepackage[
backend=biber,
style=bwl-FU,
]{biblatex}
\usepackage{textcomp}
\usepackage{hyperref}
\begin{document}
\title{Executing a Successful Third Shot Drop in Pickleball}
\author[1,2]{Douw Steyn}
\author[2]{Troy Mithrush}
\author[2]{Chris Koentges}
\author[2]{Susan Andrews}
\author[2,3]{Andre Plourde}
\affil[1]{Department of Earth, Ocean and Atmospheric Sciences\\The University of British Columbia\\Vancouver, B.C., Canada \\{\textit{dsteyn@eoas.ubc.ca}}}
\affil[2]{Jericho Hill Pickleball School\\Vancouver, B.C., Canada\\{\url{https://jerichohillpickleball.com}}}
\affil[3]{Department of Economics\\Carleton University\\Ottawa, Ontario, Canada}
\date{\today}

	\maketitle
	\newpage
	\begin{abstract}
		
	We define and investigate a successful third shot drop in pickleball using a numerical model of pickleball ball aerodynamics. Our overall objective is to investigate the ranges of initial speeds, angles and spins that result in a successful third shot drop. We conclude that the initial speed must be in the range $10.9~ms^{-1}$ to $13~ms^{-1}$ for down-the-line shots and $13.3~ ms^{-1}$ to $16~ms^{-1}$ for cross-court shots.  The initial angle must be in the range $15.5$ degrees to $22.5$ degrees for down-the-line shots and $12.5$ degrees to $18$ degrees for cross-court shots.  We conclude that the effects of spin on the third shot drop are of secondary importance. We believe these results could be useful as a guide to coaches and players wanting to develop this crucial aspect of the game of pickleball.
		
	\end{abstract}
	
	\section{Introduction}
	\label{sec:intro}

 Pickleball is reportedly the fastest growing sport in North America as reported by \cite{SciAm}, and is increasingly being played internationally.  The popular press has taken notice, for example:  \cite{DOGONEWS}  and \cite{GUARDIAN}, and there is a flood of online videos illustrating the game, as played by both professionals and amateurs.  Reasons for the popularity of pickleball include: low levels of athleticism required for entry-level play, relatively light costs of equipment and clothing, high levels of sociability among amateur players, and a ball game that is filled with fast-paced action and high levels of excitement. \cite{Playpball} on the \emph{Play Pickleball} web site suggests that pickleball could be introduced to the Olympics for the 2032 summer games. 
 
 Pickleball is a "racket and ball" game played in either singles or doubles on a court much smaller than a tennis court - 5.2 $m$ (20 feet) wide with baseline to net distance of 6.71 $m$ (22 feet).  The net is slightly lower than a tennis net - 0.86 $m$ (34 inches) at the sidelines and 0.81 $m$ (32 inches) at mid-court.  Without loss of generality, this paper is structured around a doubles match. The racket, called a paddle in pickleball, has a rigid face and the ball is of low-bounce hard plastic with between 26  and 40 holes for indoor and outdoor play respectively.  An important feature of the pickleball court is the \emph{Non-Volley Zone}, commonly called the kitchen, which is 2.13 $m$ (7 feet) from the net.  As the name implies, the ball cannot be volleyed while the player is inside this zone. 
 
 The main objective of this study is to present useful advice to both players and coaches regarding the characteristics (initial speed, angle and spin) of a played stroke that leads to a successful third shot drop (after the serve and return-of-serve).  To achieve this, we develop a mathematical model of the flight (trajectory) of a pickleball and use the model to investigate the third shot drop.
	
	\subsection{The Game of Pickleball}
	\label{sec:game}
 
 In an interesting deviation from related racket and ball games played on a court, the rules of pickleball are designed to reduce the power of the serve, and to eliminate the possibility of the serve and volley sequence so common in tennis. This is achieved by requiring that the serve be struck underhand from behind the baseline, and must bounce between the kitchen line and baseline. In addition, the return-of-serve must be struck only after a bounce. As a result, the third shot in a rally is often struck near the baseline.  The third shot is thus the first shot that can be played in an unrestricted way. A common strategy for the third shot is to play it as a drop shot that lands inside the opposing kitchen.  This is called a \emph{third shot drop}.  It is a shot often practised in drills by beginner players, and almost universally used in competitive play at all levels. The third shot drop is a much-discussed aspect of pickleball play, and is the subject of a multitude of freely-available training videos.  A prominent example is given by \cite{TSDvideo}.  As in the majority of similar videos, the author makes recommendations about player positioning after the serve, paddle grip and swing, and very general advice about the desired ball trajectory.  A universal piece of advice is to achieve a trajectory that reaches an apex before crossing the net, and to aim for a bounce in the far part of the opposing kitchen. The conventional wisdom among pickleball players is that doing so reduces the options available to opposing players when playing the fourth shot of the rally.  These features of the third shot drop are well explained by \cite{dummies}, who also outline general stroke techniques for making the shot.

 Because the game of pickleball is so recent, very little work has been done on technical aspects of the play, and specifically not much is known about the aerodynamic characteristics of the pickleball ball.  This will require us to start our work by an analysis of pickleball ball flight so as to determine the two main aerodynamic characteristics of ballistic objects: the drag and lift coefficients.
	
\subsection{Modelling Pickleball Trajectories}
\label{sec:flight}

 There exists a wide and varied scientific literature on the aerodynamics of sports balls.  An early general review of the aerodynamics of balls in a wide range of sports was provided by \cite{Mehta}. 
 
 Specific sport balls have been studied by \cite{soccer} for soccer balls, \cite{baseball} for baseball, by \cite{BandH1976} and \cite{golf} for golf, and \cite{CrossLindsey2017} and \cite{ivanov} for tennis. While many studies have examined 2-D ball trajectories, \cite{ivanov} and \cite{golf} are notable in that they have examined 3-D trajectories.  Most studies of ball flight in sports use aerodynamic engineering methods, combined with force-balance numerical models.  \cite{wen} provides a notable recent advance in the use of deep-learning or Artificial Neural Network methods to analyze baseball trajectories.
 
Explicit modeling of the aerodynamics of a pickleball has been carried out by \cite{creer2023computational}, who uses computational fluid dynamics modelling in an attempt to estimate lift and drag coefficients of 26- and 40-hole pickleballs. The work is inconclusive because of computational difficulties. More relevant to our present work is the study of \cite{emond2024pickleball} who models pickleball flight trajectories in order to investigate the advantage or disadvantage of upwind versus downwind play.  This work does not consider lift, and uses a drag coefficient of $C_d=0.6$ based on measured drag coefficients of a wiffleball. The work emphasizes the need for measurements to provide a direct estimate of  $C_d$ . 
 
 The majority of these studies present conclusions that are only useful to players and coaches with unusually deep technical backgrounds.  Our objective is to conduct modelling of the flight of a pickleball and provide an interpretation of the results that will be useful for players and coaches of the game. Our interests thus go beyond ball aerodynamics to focus on a strategy for executing a successful third shot drop.
 
 Our development of the pickleball flight model, and its use to determine drag and lift coefficients for a pickleball ball are shown in Appendix \ref{sec:drag_lift}.  Our model is based on well-known equations for the aerodynamics of flying, spinning objects.  The measured trajectories of a pickleball in flight are used to determine the two characteristics that quantify aerodynamic lift and drag (the lift and drag coefficients respectively). This model is then recast  in an exploratory mode to investigate the dependence of the ball trajectory on the three strike parameters: initial speed, strike angle and spin rate.  The initial conditions will represent a typical game context in which the third shot is struck from the baseline $x=0.0~m$ and at roughly knee height $z=0.75 ~m$. Our requirement for the third shot drop is that the ball clears the net, and bounces somewhere within the kitchen. In the exploratory model, the three strike parameters are specified as run conditions.  As described in Appendix \ref{sec:drag_lift}, the spin rate is used to determine the lift coefficient through the spin parameter.

\begin{figure}[H]
  \begin{center}
		\includegraphics[width=0.8\textwidth]{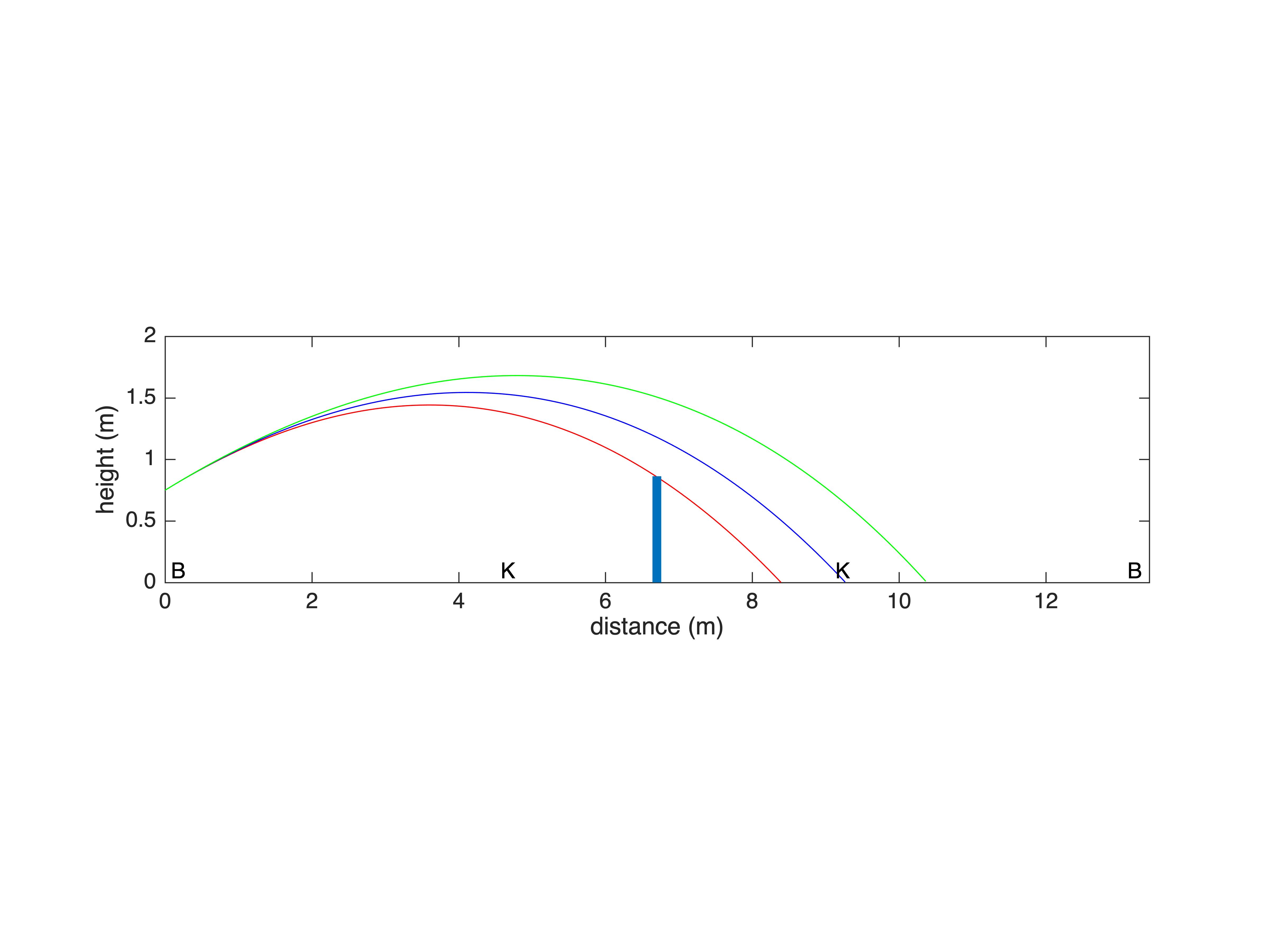}
		\captionof{figure}{Three modelled pickleball trajectories with backspin ($C_l=-0.12$, green trajectory), no spin ($C_l=0.0$, blue trajectory) and topspin ($C_l=+0.12$, red trajectory). The blue vertical bar depicts the net, "K" indicate kitchen lines, and "B" indicate baselines.  In these trajectories, $C_d = 0.30$, initial speed $= 12~ms^{-1}$, initial angle $=20^{\circ}$, initial spin rate $= 10.0 s^{-1}$, spin parameter $=0.19$. Reynolds number ranged from $5.3\times10^4$ to $6.4\times10^4$.}
		\label{fig:fig2}
 \end{center}
\end{figure}

 In Figure~\ref{fig:fig2} we show three shots from the model in exploratory mode, with different initial spin rates. These shots illustrate the overall model performance, and show the effect of varying spin.  All three shots had the same spin rate of $10~s^{-1}$  but the sign (positive or negative) of the resulting lift coefficient was set to reflect the spin sense (backspin or topspin). As is evident from Figure \ref{fig:fig2} the model captures a strong spin effect on ball trajectory.  

The following section will explore the idea of a successful third shot drop, and then use the model to understand how such a shot may be achieved through combinations of initial speed, strike angle and spin rate.  

\subsection{The Third Shot Drop}
\label{sec:drop}

\begin{figure}[H]
\begin{center}
		\includegraphics[width=0.6\textwidth]{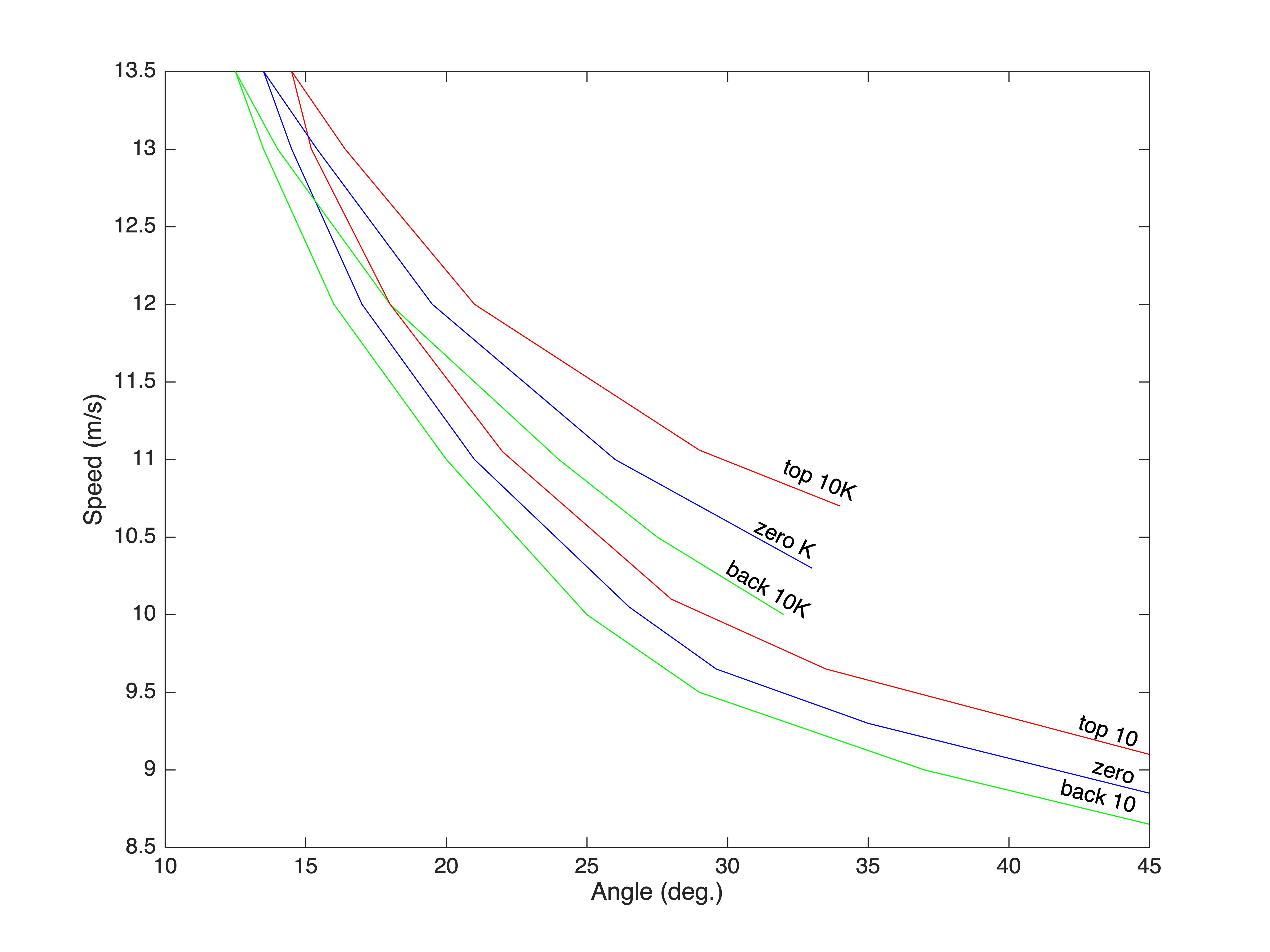}
		\caption{Combinations of angle, speed and spin rate that result in third shot drops that land in the kitchen. "top" (red lines) and "back" (green lines) indicate topspin and backspin shots respectively. "K" indicates shots that land on the kitchen line. We plot only spin rates of $10~s^{-1}$, as lower spin rate shots will fall between the respective lines. Blue lines are for zero spin rate.}
		\label{fig:angleandspeed}
  \end{center}
	\end{figure}

Repeated running of the trajectory model with specified shot speed, angle and spin rate can be used to investigate the third shot drop. It will always be possible to achieve a given total range with two possible shots, one having a high (generally greater than 45 degrees) shot angle and the other a much lower shot angle.  The high angle shot is of no interest since such shots will give opposing players a lot of time to react. Furthermore, a high angle shot will result in a high bounce, giving the opposing players an opportunity to attack the ball.  We therefore only consider the low-angle shots.  There exist two possible limiting shots: one that just clears the net and bounces somewhere in the kitchen, and the other that clears the net widely and bounces exactly on the kitchen line.  Figure~\ref{fig:angleandspeed} shows the result of this exploration for shots made at $0.75~m$ high, on the baseline.  A number of features are worth noting:
\begin{itemize}
    \item At speeds below about $8.7~ms^{-1}$ it is not possible to clear the net with shots from the baseline.
    \item At speeds higher than about $13.5~ms^{-1}$ the ball will always land beyond the kitchen line.
    \item Topspin shots with a spin rate of $10~s^{-1}$ and strike angles greater than 35 degrees will not clear the net. For backspin strikes, the limiting angle is 33 degrees.
    \item The convergent lines at $13.5~ms^{-1}$ indicate trajectories of balls that bounce on the kitchen line. 
    \item Shots that fall in the space between "top10K" and "back10" can be considered legitimate third shot drops only in the sense that they land between net and kitchen line.  In the coming analysis we will narrow our consideration of all third shot drops to consider only those that can be considered successful.
    \item We plot lines only for spin rates of $10~s^{-1}$ topspin and backspin and zero spin.  Spin rates lower than $10~s^{-1}$ fall at intermediate positions and make for a very crowded figure. Spin rates up to $15~s^{-1}$ are noted as possible by \cite{pballspin}, limited only by ball and paddle roughness. We judge such extreme spin rates to be unlikely in most play by recreational pickleball players. 
\end{itemize}

\begin{figure}[H]

\begin{center}
		\includegraphics[width=0.9\textwidth]{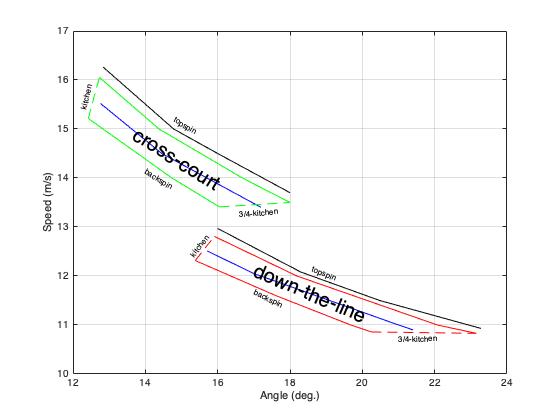}
		\caption{Combinations of angle, speed and spin rate that result in third shot drops that would be strategically strong, landing near the back one quarter of the kitchen. Red/green lines define ranges of speed and angle for down-the-line/cross-court shots with spin rates of $10~s^{-1}$. Lower spin rate shots will fall between the respective topspin and backspin lines. Black lines show shots with topspin of $15~s^{-1}$, blue lines show shots with no spin.}
		\label{fig:angleandspeed2}
  \end{center}
	\end{figure}

Figure ~\ref{fig:angleandspeed} depicts all possible low-angle shots that clear the net and bounce in the kitchen with down-the-line (parallel to court sidelines) trajectories.  As discussed, the strategically strongest third shot drops are ones that bounce somewhere close to the kitchen line in order to prevent the serve receivers from making the fourth shot an attack shot. For convenience we model third shot drops that bounce beyond three quarters way between net and kitchen line, but short of the kitchen line. We also model only shots that barely clear the net.  This is to ensure that the ball's (downward) vertical velocity is a minimum possible, and so the bounce height of the ball is also a minimum. These conditions are consistent with the general guidance given by \cite{TSDvideo} and \cite{dummies}.  It is also important to consider cross-court third shot drops.  For these reasons we develop a figure that is similar to, but a subset of Figure ~\ref{fig:angleandspeed}, and include both down-the-line and cross-court shots.  For completeness, we include topspin shots with spin rates of $15~s^{-1}$, and assume that backspin rates this high will be difficult to achieve. These are shown in Figure~\ref{fig:angleandspeed2}.

 An examination of Figure \ref{fig:angleandspeed2} reveals the following features that are relevant for players making and developing their skills at executing a third shot drop, and for coaches instructing players on the shot:

 \begin{itemize}
   \item The range of ball speeds for a successful third shot drop is $10.9~ms^{-1}$ to $13~ms^{-1}$ for down-the-line shots and $13.3~ ms^{-1}$ to $16~ms^{-1}$ for cross-court shots. These are relatively narrow speed ranges, and while it is unlikely even expert players will be able to calibrate their shot speed with this precision, it does suggest that developing skill in shot speed control is very important in making a successful third shot drop.
    \item The range of shot angles for a successful third shot drop is $15.5$ degrees to $22.5$ degrees for down-the-line shots and $12.5$ degrees to $18$ degrees for cross-court shots.  As with shot speed, these are relatively narrow shot vertical angles, and suggest developing skill in shot angle control is very important in making a successful third shot drop.
    \item The effects of varying speed and angle are much stronger than the effect of varying spin. This is evident in Figure \ref{fig:angleandspeed2} from the relatively small differences in speed and angle that result from extreme variation between topspin and backspin. Spin variation between a topspin of $15 ~s^{-1}$ and backspin of $10~s^{-1}$ produce a trajectory variation less than that produced by a speed variation of about $0.8~ms^{-1}$ for down-the-line shots and roughly $1.0~ms^{-1}$ for cross-court shots. These observations suggest that players wanting to execute successful third shot drops should concentrate on angle and speed control, rather than on spin development.
\end{itemize}

	\section{Conclusions} \label{sec:conclusion}

 Expert pickleball players and coaches agree that the trajectory of a successful third shot drop in pickleball reaches its apex before crossing the net, and bounces in the far portion of the opposing kitchen.  These characteristics give the serving side the maximum chance of continuing in a successful rally. 
 
 We investigated the third shot drop in pickleball, using a numerical model of pickleball flight.  The drag and lift coefficients of a 40-hole pickleball were determined by fitting modelled to measured pickleball trajectories, as detailed in Appendix \ref{sec:drag_lift}.  Using the model in an exploratory mode, we were able to determine ranges of initial speed, angle and spin that resulted in trajectories producing a successful third shot drop. 
 
 We conclude that:
 \begin {itemize}
 \item The initial speed must be in the range $10.9~ms^{-1}$ to $13~ms^{-1}$ for down-the-line shots and $13.3~ ms^{-1}$ to $16~ms^{-1}$ for cross-court shots.  
 \item The initial angle must be in the range $15.5$ degrees to $22.5$ degrees for down-the-line shots and $12.5$ degrees to $18$ degrees for cross-court shots.  
 \item The effects of spin on the third shot drop are of secondary importance in determining where the ball bounces. We acknowledge that ball spin can have a strong effect on the post-bounce trajectory. 
 \end{itemize}
 
 We acknowledge that speed and angle control at the level of precision noted above can only be achieved with considerable practice and skill. Using immediate playback video recording of shots in practice sessions to determine initial speed and angle could be a beneficial coaching tool to achieve such shot control.
	
	\paragraph{Acknowledgements} 
	
The study was self-funded by the authors.  All authors contributed to this study, although in different ways, and assume responsibility for the conclusions. The authors report no conflict of interest.  
	
       \begin{appendix}
       
       \section{Developing a Pickleball Trajectory Model} \label{sec:drag_lift}

 We develop a simple, two-dimensional (vertical and horizontal coordinates only) numerical model of a pickleball trajectory, assuming that initial conditions of speed, angle and spin are set by the player's swing. Fixed parameters in the model must include physical characteristics of the pickleball ball.  Our model is essentially the same as that used by \cite{emond2024pickleball}, the only difference being in the numerical implementation.  The official ball has  a diameter between 2.874 inches (0.0730 m) and 2.972 inches (0.0755 m) and a mass between 0.78 oz (0.0221 kg) to 0.935 oz (0.0265 kg). For simplicity we use averages of the limiting values.  

As noted, the pickleball ball is unusual in that it is perforated by between 26 and 40 holes.  This means that, at least at low Reynolds numbers, secondary flows inside the ball are possibile.  This will vastly complicate the aerodynamics of the ball.  At high Reynolds numbers, it is likely that the pickleball ball simply behaves as a rough ball.  
 
 \cite{pballsci} shows that, in order to clear the net and land near the back of the court, a pickleball serve must have an initial speed between 18 and $24~ms^{-1}$.  It is likely that the third shot will be well below this range of speeds since the third shot drop should land the ball inside the opposing kitchen.  We will show that the Reynolds number will be in the range $5\times10^4$ to $5\times10^5$, and will therefore treat a pickleball as if it were a solid, rough ball.

	\subsection{Model Equations}
	
	Once struck, a ball in flight follows a trajectory governed by Newton's second law:
	
	\begin{equation}\label{eqn:forces}
	Acceleration =  \frac{F_{drag} + F_{lift} + F_{gravity}}{mass}.
	\end{equation}

	While paddle-induced ball spin in pickleball is generally on an off-horizontal axis, we restrict ourselves to spin only on a horizontal axis, perpendicular to the line of flight.  This leads to the great simplification of only two-dimensional (in the $x-z$ plane) trajectories, and yet captures the major spin effects in a real pickleball game.

	Drag force is given by: $F_{drag}={C_d}\rho_{a}{A}{u^2}$, where $C_d$ is the drag coefficient, $\rho_{a}$ is air density, $u^2$ is the square of ball speed, and $A$ is ball frontal area. $F_{lift} $ has a similar form, except that $C_l$ is the lift coefficient.

Expanding equation \ref{eqn:forces} in $x$ and $z$ component form gives:

\begin{equation}\label{eqn:zforce}
\frac{{d{^2}}{z}}{dt{^2}} = -  \frac{\rho_{a}{A}}{2m}[v^2_z+v^2_x]\left({C_d}{\sin{\theta}} - {C_l}{\cos}{\theta}\right) - g 
\end{equation} 
  
\begin{equation} \label{eqn:xforce} 
\frac{{d{^2}}{x}}{dt{^2}} =  -  \frac{\rho_{a}{A}}{2m}[v^2_z+v^2_x]\left({C_d}{\cos{\theta}} - {C_l}{\sin}{\theta}\right),
\end{equation}    
where $v_x$ and $v_y$ are $x$ and $z$ components of ball velocity, respectively, and $\theta$ is instantaneous angle of ball trajectory, and $\rho_{a}$ is air density, and $m$ is the ball mass.
 
 Since $v_x$ and $v_y$ and $\theta$ are all time dependent, equations \ref{eqn:zforce} and \ref{eqn:xforce} are not integrable.  We employ a simple Euler scheme to step equations \ref{eqn:zforce} and \ref{eqn:xforce} forward in time, from specified initial conditions for height, angle, spin and speed.

	\subsection{Determining Drag- and Lift-Coefficients of a Pickleball}
	
\cite{BandH1976} determined the drag ($C_d$) and lift ($C_l$) coefficients of sports balls in flight by laboratory (often wind-tunnel) measurements, while \cite{CrossLindsey2017} and \cite{wen} determined $C_d$ and $C_l$ from full-scale balls in flight using one or more video cameras. As noted, \cite{emond2024pickleball} assumed a value of $C_d=0.0$ based on the similarity between a pickleball and a wiffleball.  In our work, we chose to film the trajectory of a 40 hole \textregistered Selkirk pickleball in flight from a single camera and to use the digitized trajectories to determine the two aerodynamic coefficients.  The single camera approach is possible if camera and trajectory are set up to allow calculation of $x$ and $z$ coordinates of the ball in flight using simple projective geometry.  The requirements are that the projection of the ball trajectory on the court floor is known.  This is achieved by ensuring that the ball flight is directly above one of the court sidelines.  Reference vertical dimension is provided by the net height at the ball trajectory.  Reference horizontal dimension is provided by baseline, net and kitchen lines, again at the ball trajectory.  A further simplification is provided by setting the camera at the net height.   Menelau's projection theorem is needed to determine the $x$ coordinate from measurements of ball position in the frame.

The camera gave individual frames every 0.008 $s$ and we used every $10^{th}$ frame to yield roughly 12 to 20 digitized points in every trajectory. We used the first two points of each trajectory to determine initial speed and angle of the shot, and initialized our model at the first digitized point.  Spin rate was determined by marking the ball with three orthogonal great circles, and estimating the number of revolutions between roughly every 5 to 10 frames.  Given initial speed, initial angle, initial coordinates and spin rate, the only two remaining variables that define an individual trajectory are $C_d$ and $C_l$.  These two quantities were then estimated by minimizing deviation between calculated trajectory and digitized trajectory points.

\begin{figure}[H]

		\centering
		\includegraphics[width=0.95\textwidth]{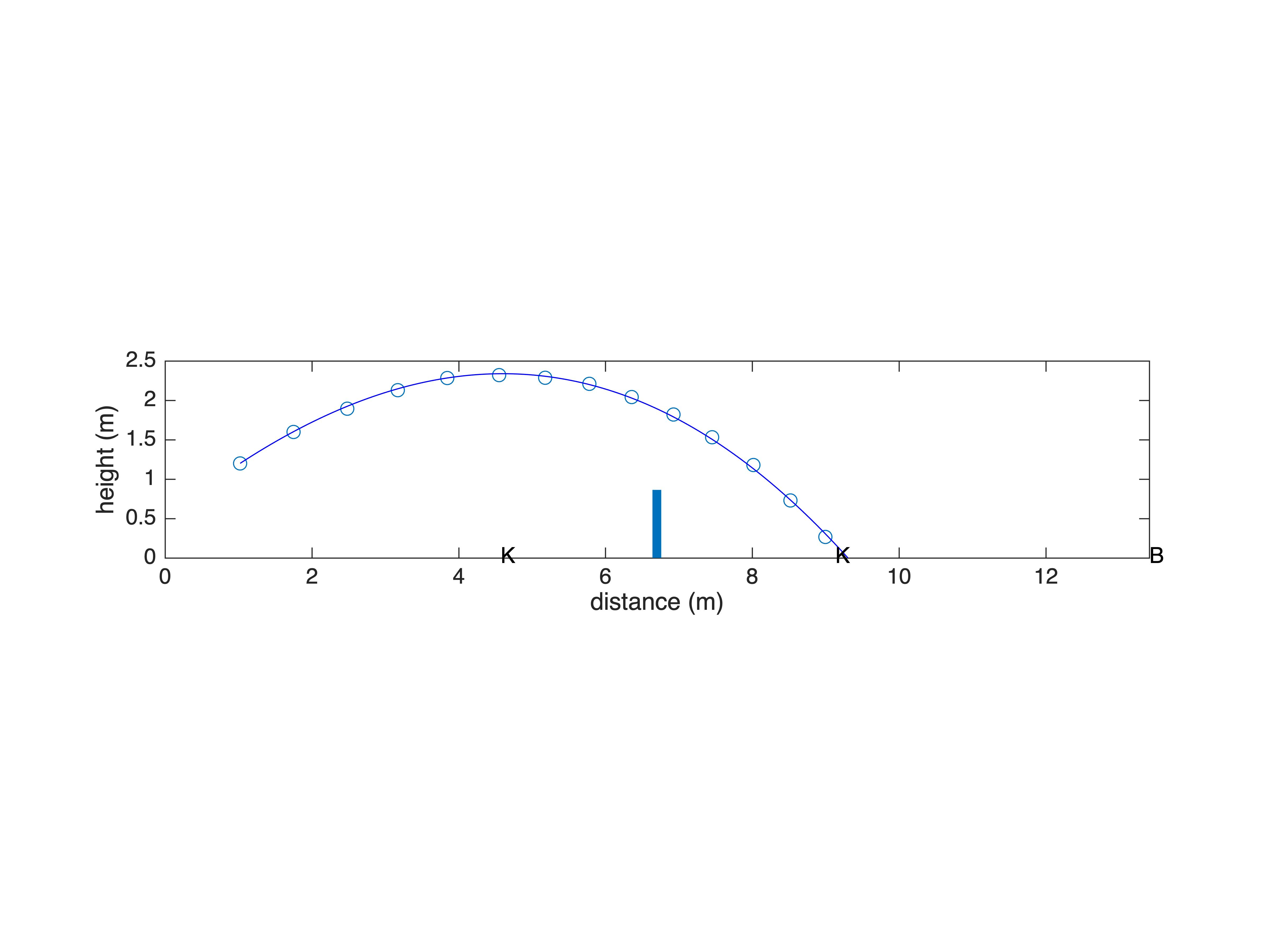}
		\caption{Measured (points) and modelled (line) pickleball trajectory. The blue bar depicts the net, "K" indicate kitchen lines, and "B" indicates the far baseline. The x axis origin is near the baseline.  Data points are roughly the size of a pickleball. In this trajectory, the initial speed was $10.0~ms^{-1}$; the initial angle was $31^{\circ}$; the Reynolds number ranged from $5.4\times10^4$ to $4.9\times10^4$, spin rate was 5.9 $s^{-1}$, giving a spin parameter of $0.14$ and a time of flight of  $1.42$ sec.}
		\label{fig:modelled_trajectory}
	\end{figure}

Figure~\ref{fig:modelled_trajectory} shows one of the six measured and modelled trajectories.  In all measured trajectories, the modelled trajectory matched the measured trajectory to the same degree as in the illustrated trajectory.

The error in determining the trajectory points analysis arises from error in measurement of ball position on the individual frames.  A simple error analysis indicates that the total error should not be greater than twice the ball diameter.  Given that these errors should be random, and that we fit a trajectory over 10 to 20 points, the resultant error in $C_l$ and $C_d$ should be very small.  A further source of error is due to the ball trajectory not being exactly over the presumed projection line (court sideline).  It is extremely difficult to precisely quantify this error.
\par
In their dimensional analysis of golf ball flight, \cite{BandH1976} give $C_l$ and $C_d$ as:

\begin{equation}
C_l, C_d = F(R_e, S),
\end{equation} 
where the Reynolds number $R_e=\frac{2{u}\rho_{a}}{\nu_{a}}$ and the spin parameter $S=\frac{2\pi{r}{n}}{u}$. The rotation speed $n$ is in $s^{-1}$, $r$ is the ball radius in $m$ and $u$ is the ball speed in $ms^{-1}$. Air density $\rho_{a} = 1.29 kg{m^{-3}}$ and air viscosity $\nu_{a} = 1.78 kg{m^{-1}}s^{-2}$. $F$ is a function to be determined by measurement.

In our six measured shots, the ranges of the two independent dimensionless parameters are: $4.7\times10^4<R_e< 6.1\times10^4$ for the initial $R_e$ and  $0.03<S<0.3$.  These ranges are small relative to parameter ranges  found for golf by \cite{BandH1976} and tennis by \cite{CrossLindsey2017}, but are likely typical of the parameter values for pickleball.  Our data show no systematic dependence of $C_d$ on $S$ and $R_e$, the average is:

\begin{equation}
  C_d = 0.30\pm0.02.  
\end{equation}

The constancy of our value of $C_d$ (over the range of $R_e$) is consistent with the assumption of \cite{emond2024pickleball}, but our absolute value is significantly lower. 

Our data show an increase of  $C_l$ with $S$, with large scatter, likely due to uncertainty in measured spin rate.  The fitted linear dependence, but with forcing to zero $C_l$ for zero spin is:

\begin{equation}
 C_l = 0.195S.  
\end{equation}

This form has roughly the same linear dependence on $S$ as shown in Figure 3 of \cite{CrossLindsey2017}. We believe these are the first determinations of $C_d$ and $C_l$ and their $S$ dependence for a pickleball ball. It would be beneficial to confirm our values of  $C_d$ and $C_l$ in a wind tunnel.  The question of dependence of  $C_d$ and $C_l$  on the number of holes remains open.

We use the average values of $C_d$, and the $S$ dependent value of $C_l$  in our exploratory modelling of pickleball third shot drop trajectories in sections \ref{sec:flight} and \ref{sec:drop}.

\end{appendix}

\printbibliography	
	
\end{document}